\documentclass[%
reprint,
superscriptaddress,
amsmath,amssymb,
aps,
prl,
]%
{revtex4-2}

\usepackage[utf8]{inputenc}
\usepackage[T1]{fontenc}
\usepackage[english]{babel}
\usepackage[autostyle=true]{csquotes}

\usepackage{xcolor}
\usepackage{graphicx}
\graphicspath{{figures/}}
\usepackage{dcolumn}
\usepackage{bm}
\usepackage{hyperref}
\hypersetup{
	    pdftitle={Exciting the Goldstone Modes of a Supersolid Spin-Orbit-Coupled Bose Gas},
	    pdfauthor={Kevin T. Geier, Giovanni I. Martone, Philipp Hauke, Sandro Stringari},
		colorlinks=true,
		allcolors=blue,
}

\usepackage[%
	caption=false,
	subrefformat=simple,
	labelformat=simple,
]{subfig}

\usepackage[%
	separate-uncertainty=false,
]{siunitx}
\DeclareSIUnit\gauss{G}

\usepackage{braket}

\usepackage[capitalise]{cleveref}

\newcommand{\vect}[1]{\bm{#1}}
\newcommand{\diff}{\mathop{}\!\mathrm{d}}
\newcommand{\recoilenergy}{\ensuremath{E_{\mathrm{r}}}}
\newcommand{\omegacritical}{\ensuremath{\Omega_{\mathrm{cr}}}}
\newcommand{\upcomponent}{\ensuremath{\uparrow}}
\newcommand{\downcomponent}{\ensuremath{\downarrow}}
\newcommand{\densitycomponent}{\ensuremath{\mathrm{n}}}
\newcommand{\spincomponent}{\ensuremath{\mathrm{s}}}

\begin{document}

\title{Exciting the Goldstone Modes of a Supersolid Spin--Orbit-Coupled Bose Gas}
\date{\today}

\author{Kevin T. Geier}
\email{kevinthomas.geier@unitn.it}
\affiliation{INO-CNR BEC Center and Dipartimento di Fisica, Universit\`a di Trento, 38123 Povo, Italy}
\affiliation{Institute for Theoretical Physics, Ruprecht-Karls-Universität Heidelberg, Philosophenweg 16, 69120 Heidelberg, Germany}
\author{Giovanni I. Martone}
\affiliation{Laboratoire Kastler Brossel, Sorbonne Universit\'{e}, CNRS, ENS-PSL Research University, Coll\`{e}ge de France; 4 Place Jussieu, 75005 Paris, France}
\author{Philipp Hauke}
\affiliation{INO-CNR BEC Center and Dipartimento di Fisica, Universit\`a di Trento, 38123 Povo, Italy}
\author{Sandro Stringari}
\affiliation{INO-CNR BEC Center and Dipartimento di Fisica, Universit\`a di Trento, 38123 Povo, Italy}

\begin{abstract}%
Supersolidity is deeply connected with the emergence of Goldstone modes, reflecting the spontaneous breaking of both phase and translational symmetry. Here, we propose accessible signatures of these modes in harmonically trapped  spin--orbit-coupled Bose--Einstein condensates, where supersolidity appears in the form of stripes. By suddenly changing the trapping frequency, an axial breathing oscillation is generated, whose behavior changes drastically at the critical Raman coupling. Above the transition, a single mode of hybridized density and spin nature is excited, while below it, we predict a beating effect signaling the excitation of a Goldstone spin-dipole mode. We further provide evidence for the Goldstone mode associated with the translational motion of stripes. Our results open up new perspectives for probing supersolid properties in experimentally relevant configurations with both symmetric as well as highly asymmetric intraspecies interactions.%
\end{abstract}

\maketitle

Supersolidity is an exotic state of matter characterized by the simultaneous spontaneous breaking of $U(1)$ symmetry, yielding superfluidity, and of translational invariance, yielding crystallization~\cite{Thouless1969,Andreev1969,Leggett1970}. In the past, this state of matter has attracted considerable attention in the context of solid helium, where the experimental efforts to reveal the effects of superfluidity, however, have not been conclusive~\cite{Balibar2010,Boninsegni2012}. More recently, a renewed interest has emerged in the context of ultracold atomic gases, and experimental evidence of typical supersolid features has been reported in Bose--Einstein condensates inside optical resonators~\cite{Leonard2017}, spin--orbit-coupled configurations~\cite{Li2017,Putra2020}, and in cold gases interacting with long-range dipole forces~\cite{Tanzi2019,Boettcher2019,Chomaz2019}.
An important property of supersolidity with respect to ordinary superfluids is the appearance of additional gapless (Goldstone) modes in the excitation spectrum, resulting from the broken translational symmetry~\cite{Andreev1969,Josserand2007,Saccani2012,Kunimi2012,Li2013,Macri2013,Liao2018,Roccuzzo2019,Lyu2020,Martone2021,Hofmann2021}.
While these modes have already been the object of first experimental investigation inside optical resonators~\cite{Leonard2017a} and in harmonically trapped dipolar gases~\cite{Tanzi2019b,Guo2019,Natale2019,Petter2021}, so far no experimental observation has been reported in spin--orbit-coupled configurations.

An important motivation behind the present Letter is to better understand the significance of the spin degree of freedom for the excitation mechanisms of the Goldstone modes in the supersolid phase. Focusing on a harmonically trapped spin--orbit-coupled Bose--Einstein-condensed gas, we identify a characteristic beating effect that allows for the observation of a Goldstone excitation of spin nature, in analogy to a similar procedure followed in dipolar supersolids~\cite{Tanzi2019b}.
Moreover, we show that the locking of the polarization after a uniform spin perturbation provides evidence for the Goldstone mode associated with the rigid translation of the stripes~\cite{Guo2019}.

Spin--orbit-coupled Bose--Einstein condensates are known to exhibit an intriguing variety of quantum phases, which are obtained by tuning the experimentally controllable Raman coupling, responsible for coherence between atoms occupying two different hyperfine states~\cite{Dalibard2011,Galitski2013,Aidelsburger2018}. Spin--orbit coupling gives rise to an effective single-particle Hamiltonian of the form~\cite{Lin2011}
\begin{equation}
H_{\mathrm{SP}} = \frac{1}{2m} \left( \vect{p} - \hbar \vect{k}_0 \sigma_z \right)^2 + \frac{\Omega}{2}\sigma_x + \frac{\delta}{2} \sigma_z + V(\vect{r}) ,
\label{HSOC}
\end{equation}
where $m$ is the atomic mass, and $\sigma_x$ and $\sigma_z$ are Pauli matrices.
The spin--orbit term, fixed by the momentum transfer $\hbar \vect{k}_0 = \hbar k_0 \hat{\vect{e}}_x$ between two intersecting laser fields generating the Raman coupling with strength~$\Omega$ and effective detuning~$\delta$, is at the origin of non-trivial many-body effects which deeply differ from the ones caused by simple coherent coupling with negligible momentum transfer, such as radio frequency or microwave coupling.
We are, in particular, interested in the observability of the Goldstone modes in the presence of a harmonic trapping potential $V(\vect{r}) = m (\omega_x^2 x^2 + \omega_y^2 y^2 + \omega_z^2 z^2) / 2$, with frequencies
$\omega_i$ and associated oscillator lengths $a_i = \sqrt{\hbar / m \omega_i}$, $i = x, y, z$.

In what follows, we assume that the two-body interaction between the atoms is described within the usual mean-field Gross--Pitaevskii theory of quantum mixtures~\cite{Pitaevskii2016}. Then, writing the order parameter in the spinor form $\Psi = (\Psi_\upcomponent, \Psi_\downcomponent)^T$ with the wave functions $\Psi_\upcomponent$ and $\Psi_\downcomponent$ describing the relevant hyperfine states, the energy of the system is
\begin{equation}
E= \int \diff \vect{r} \left( \Psi^\dagger H_{\mathrm{SP}} \Psi + \frac{g_{\densitycomponent \densitycomponent} n^2}{2} + \frac{g_{\spincomponent \spincomponent} s_z^2}{2} + g_{\densitycomponent \spincomponent} n s_z \right) .
\label{E}
\end{equation}
Here, $g_{\densitycomponent \densitycomponent} = (g_{\upcomponent \upcomponent} + g_{\downcomponent \downcomponent} + 2 g_{\upcomponent \downcomponent})/4$, $g_{\spincomponent \spincomponent} = (g_{\upcomponent \upcomponent} + g_{\downcomponent \downcomponent} - 2 g_{\upcomponent \downcomponent})/4$, and $g_{\densitycomponent \spincomponent} = (g_{\upcomponent \upcomponent} - g_{\downcomponent \downcomponent})/4$ denote, respectively, the density--density, spin--spin, and density--spin interaction parameters, fixed by proper combinations of the coupling constants $g_{ij} = 4 \pi \hbar^2 a_{ij} / m$, where $a_{ij}$ are the respective scattering lengths with $i,j \in \{ \upcomponent, \downcomponent \}$.
The particle density and the spin density entering \cref{E} are defined, respectively, as $n(\vect{r})=|\Psi_\upcomponent(\vect{r})|^2+|\Psi_\downcomponent(\vect{r})|^2$ and $s_z(\vect{r})=|\Psi_\upcomponent(\vect{r})|^2-|\Psi_\downcomponent(\vect{r})|^2$, the former being normalized to the total particle number $N = \int \diff \vect{r} \, n(\vect{r})$.
Averages of an observable $O$ are defined as $\braket{O} = \int \diff \vect{r} \, \Psi^\dagger O \Psi / N$, or, with respect to an individual spin component~$i$, as $\braket{O}_i = \int \diff \vect{r} \, \Psi_i^* O \Psi_i / \int \diff \vect{r} \, |\Psi_i|^2$.
Depending on the values of the parameters entering the energy functional~\labelcref{E}, the ground state of the system can be either in the single-minimum, plane-wave, or stripe phase, hereafter also called the supersolid phase (see, e.g., Ref.~\cite{Li2015review}).
The use of the mean-field picture is justified by the fact that the quantum depletion of the condensate in these dilute quantum mixtures is a negligible effect (on the order of $\SI{1}{\percent}$), even in the presence of spin--orbit coupling~\cite{Zheng2013,Chen2018}.

In order to explore the nature of the elementary excitations, we numerically solve the coupled time-dependent Gross--Pitaevskii equations, directly derivable from the variation of the action $S = \int \diff t \, E[\Psi_\upcomponent,\Psi_\downcomponent] - i \hbar \int \diff t \diff \vect{r} \, (\Psi_\upcomponent^* \partial_t \Psi_\upcomponent + \Psi_\downcomponent^* \partial_t\Psi_\downcomponent)$ with respect to  $\Psi_\upcomponent^*$ and $\Psi_\downcomponent^*$. More specifically, we compute the ground state of the system in the presence of a static perturbation of density or spin nature  by means of a non-linear conjugate gradient method~\cite{Antoine2017}. At time $t = 0$, the perturbation is suddenly removed, and the quenched system is evolved in time using a time-splitting Fourier pseudospectral method~\cite{Antoine2013}.
The frequencies of the induced collective oscillations are extracted from sinusoidal fits to the time traces of the relevant observables.

\begin{figure}
\includegraphics[width=\columnwidth]{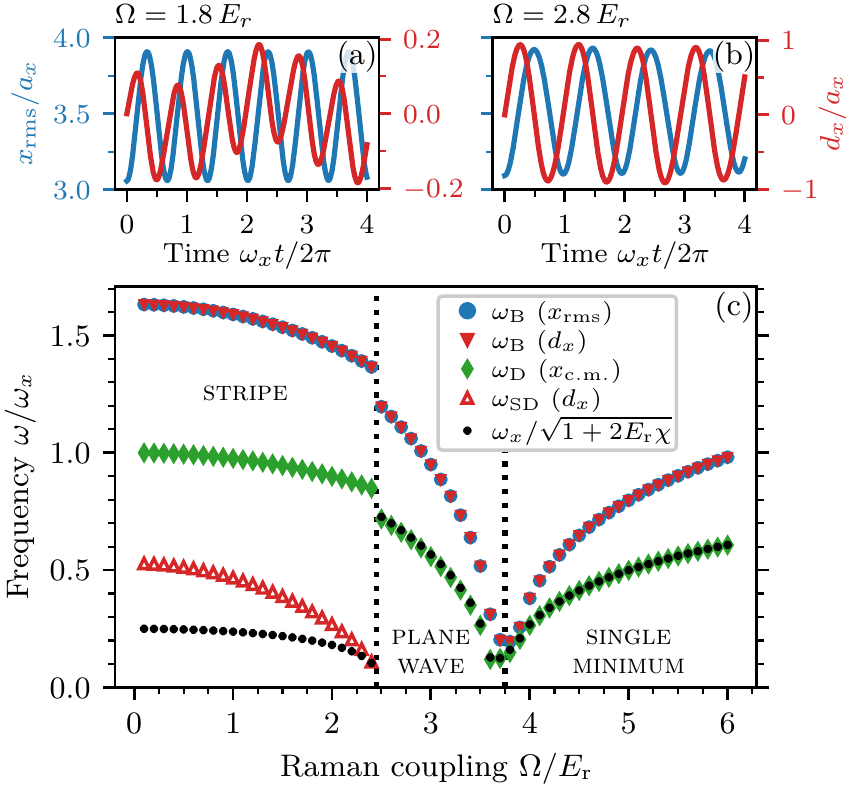}%
\subfloat{\label{fig:beating_dispersion_symmetric:stripe}}%
\subfloat{\label{fig:beating_dispersion_symmetric:pw}}%
\subfloat{\label{fig:beating_dispersion_symmetric:dispersion}}%
\caption{%
	\label{fig:beating_dispersion_symmetric}Collective modes for symmetric intraspecies interactions.
	(a),(b) Oscillations of the observables $x_\mathrm{rms} = \sqrt{\braket{x^2}}$ and $d_x = \smash[b]{\braket{x}_\upcomponent} - \smash[b]{\braket{x}_\downcomponent}$ after suddenly removing the perturbation~$H_\mathrm{pert} = \lambda m \omega_x^2 x^2$ with $\lambda = \num{0.2}$. In the stripe phase~(a), a clear beating of two frequencies $\omega_{\mathrm{B}} \approx \num{1.49} \, \omega_x$ and $\omega_{\mathrm{SD}} \approx \num{0.32} \, \omega_x$ is visible in the observable $d_x$, which is absent in the plane-wave phase~(b), where $d_x$ oscillates only at a single frequency $\omega_{\mathrm{B}} \approx \num{1.02} \, \omega_x$.
	(c) Dispersion $\omega(\Omega)$ of the breathing mode~(B), the spin-dipole mode~(SD), and the center-of-mass (dipole) mode~(D), calculated for $\lambda \ll 1$.
	The breathing and the spin-dipole modes are fully hybridized above the critical coupling~$\omegacritical \approx \num{2.5} \, \recoilenergy$, while below $\omegacritical$, a new Goldstone mode of spin nature appears.
	The dipole frequencies~$\omega_{\mathrm{D}}$ have been obtained from the center-of-mass oscillation~$x_\mathrm{c{.}m{.}} = \braket{x}$ after a sudden shift of the trap center. For $\Omega > \omegacritical$, they practically coincide with the bound $\omega_x / \sqrt{1 + 2 \recoilenergy \chi}$. The violation of this upper  bound by $\omega_{\mathrm{D}}$ for $\Omega < \omegacritical$ implies the emergence of a new low-energy mode.%
}
\end{figure}

We first investigate the case of symmetric intraspecies interactions, while the effects of strong asymmetries in the couplings are explored toward the end of this Letter.
The  majority of previous works focuses on symmetric configurations, as it is well realized, for instance, by $^{87}$Rb~\cite{Lin2011}.
For our purposes, we choose a configuration close to $^{87}$Rb with $a_{\upcomponent \upcomponent} = a_{\downcomponent \downcomponent} = \SI{100}{\bohr}$, where $\si{\bohr}$ is the Bohr radius.
Naturally, $^{87}$Rb is characterized by a value of  $g_{\upcomponent \downcomponent}$ very close to $g_{\upcomponent \upcomponent}$ and $g_{\downcomponent \downcomponent}$, yielding $g_{\spincomponent \spincomponent} \approx 0$, and hence a small value of the critical Raman coupling $\omegacritical = 4 \recoilenergy \sqrt{2 g_{\spincomponent \spincomponent} / (g_{\densitycomponent \densitycomponent} + 2 g_{\spincomponent \spincomponent})}$ for the transition to the supersolid phase~\cite{Ho2011,Li2012a}, where $\recoilenergy = \hbar^2 k_0^2 / 2 m$ is the recoil energy.
Increasing the value of $\omegacritical$ is desirable to limit the effects of magnetic fluctuations and to observe visible consequences of the presence of stripes. To this end, we consider configurations characterized by an effective reduction of the coupling constant~$g_{\upcomponent \downcomponent}$.
This may be achieved by reducing the spatial overlap between the wave functions of the two spin components, for instance, with the help of a spin-dependent trapping potential separating the two components~\cite{Martone2014,Martone2015}, or using pseudo-spin orbital states in a superlattice potential~\cite{Li2017}.
Here, we follow the former approach, using a quasi-$\mathrm{2D}$ harmonic trap with frequencies $(\omega_x, \omega_y, \omega_z) = 2\pi \times (\num{50}, \num{200}, \num{2500}) \, \si{\hertz}$, where the two spin components are separated along the $z$~direction such that the effective $\mathrm{2D}$ interspecies coupling becomes $\tilde{g}_{\upcomponent \downcomponent} = \num{0.6} \, \tilde{g}_{\upcomponent \upcomponent}$ with $\tilde{g}_{\upcomponent \upcomponent} = g_{\upcomponent \upcomponent} / \sqrt{2 \pi} a_z = \tilde{g}_{\downcomponent \downcomponent}$~\cite{Martone2014}.
The reported Raman coupling~$\Omega$ corresponds to an effective coupling, accounting for the reduced spatial overlap ~\cite{Martone2014}. Furthermore, we choose the parameters $k_0 = \sqrt{2} \pi / \lambda_{\mathrm{Raman}}$ with $\lambda_{\mathrm{Raman}} = \SI{804.1}{\nano\meter}$~\cite{Lin2011}, $N = \num{e4}$, and $\delta = 0$.

In \cref{fig:beating_dispersion_symmetric:stripe,fig:beating_dispersion_symmetric:pw}, we report the time dependence of the root-mean-square radius in the $x$~direction, $x_{\mathrm{rms}} = \sqrt{\braket{x^2}}$, and of the relative displacement of the two spin components, $d_x = \braket{x}_\upcomponent - \braket{x}_\downcomponent$, after the sudden removal of a static perturbation proportional to the operator~$x^2$, corresponding to a sudden decrease of the trapping frequency.
Since the commutator~$[x^2, H_{\mathrm{SP}}]$ contains a term proportional to $x \sigma_z$, one would expect, in general, to observe two collective oscillations of hybridized density and spin nature. For values of $\Omega$ larger than the critical coupling~$\omegacritical$, we instead find that the observables $x_{\mathrm{rms}}$ and $d_x$ oscillate with the same frequency [see \cref{fig:beating_dispersion_symmetric:pw}]. This is the consequence of the locking of the relative phase of the order parameters of the two spin components characterizing the low-frequency oscillations in the plane-wave and single-minimum phases~\cite{Martone2012}.

When we enter the stripe phase, the scenario changes drastically and we observe the appearance of a new oscillation of spin nature, revealed by the beating in the signal~$d_x$ [see \cref{fig:beating_dispersion_symmetric:stripe}]. This oscillation is the finite-size manifestation of the gapless Goldstone spin branch exhibited by the supersolid phase in uniform matter~\cite{Li2013}.
The beating reflects the fact that the two gapless branches are decoupled density and spin modes only in the limit of long wavelengths. An analogous effect has recently been identified in spin--orbit-coupled fluids of light in Kerr non-linear media~\cite{Martone2021a}. We have verified that the new mode can also be excited by applying a perturbation proportional to $x \sigma_z$, corresponding to a relative displacement of the two spin components.

In \cref{fig:beating_dispersion_symmetric:dispersion}, we report the dispersion $\omega(\Omega)$ of the resulting breathing and spin-dipole excitations.
The frequencies have been calculated for small perturbations in the regime of linear response, but the beating effect is clearly visible also for larger perturbation strengths [cf.~\cref{fig:beating_dispersion_symmetric:stripe}] that are closer to the onset of non-linearities.
Similar dispersion laws have been obtained in Ref.~\cite{Chen2017} by solving the Bogoliubov equations for a spin--orbit-coupled mixture in one dimension. With respect to Ref.~\cite{Chen2017}, our approach explicitly exposes the beating effect between the spin-dipole excitation and the compression mode in the stripe phase, as well as the full hybridization of the two modes above~$\omegacritical$.

In the limit $\Omega \to 0$, the spin-dipole frequency can be calculated analytically within the formalism of two-fluid hydrodynamics~\cite{Pitaevskii2016}. We find
\begin{equation}   
\omega_{\mathrm{SD}}^2(\Omega \to 0) = \omega_x^2\frac{1-(g_{\densitycomponent \spincomponent} / g_{\spincomponent \spincomponent})^2}{g_{\densitycomponent \densitycomponent} / g_{\spincomponent \spincomponent} - (g_{\densitycomponent \spincomponent} / g_{\spincomponent \spincomponent})^2} ,
\label{omegaSD}
\end{equation}
yielding the value $\omega_{\mathrm{SD}} = \num{0.5} \, \omega_x$ for the configuration considered here, in agreement with \cref{fig:beating_dispersion_symmetric:dispersion}.
The dispersion of the spin-dipole branch decreases as $\Omega$ approaches the transition at the critical value~$\omegacritical$ and is expected to vanish at the spinodal point, corresponding to a value of $\Omega$ a little higher than $\omegacritical$ where the system develops a dynamic instability associated with the divergent behavior of the magnetic polarizability~\cite{Li2012}.
The decrease of $\omega_{\mathrm{SD}}$ as a function of $\Omega$ is a crucial consequence of spin--orbit coupling and of the presence of stripes. By contrast, in the presence of radio frequency or microwave coupling, the spin-dipole frequency increases with the coupling strength, quickly approaching the value $\Omega$ of the spin gapped branch~\cite{Sartori2015}.

\begin{figure}
\includegraphics[width=\columnwidth]{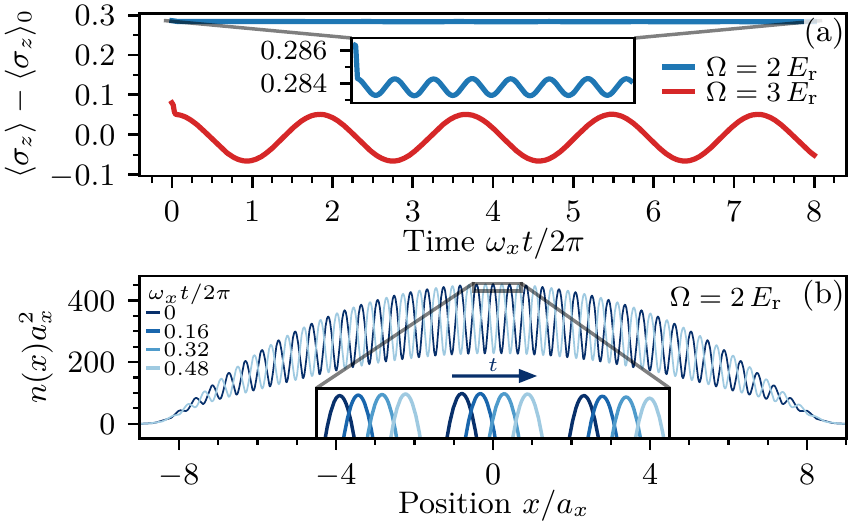}%
\subfloat{\label{fig:zero_frequency_goldstone:polarization}}%
\subfloat{\label{fig:zero_frequency_goldstone:density}}%
\caption{%
	\label{fig:polarization_symmetric}Evidence for the zero-frequency Goldstone mode associated with the translation of the stripes.
	(a) Time evolution of the polarization~$\braket{\sigma_z}$ with respect to its equilibrium value~$\braket{\sigma_z}_0$ after removing the perturbation~$H_{\mathrm{pert}} = -\lambda \recoilenergy \sigma_z$ with $\lambda = \num{0.02}$ in the stripe phase ($\Omega = 2 \, \recoilenergy$) and $\lambda = \num{0.1}$ in the plane-wave phase ($\Omega = 3 \, \recoilenergy$). In the latter case, the polarization oscillates around equilibrium at the dipole frequency~$\omega_{\mathrm{D}} \approx \num{0.55} \, \omega_x$. By contrast, in the stripe phase, the polarization remains locked for a time much longer than $2 \pi / \omega_x$, providing evidence for the zero-frequency Goldstone mode. The low-amplitude oscillations at the dipole frequency~$\omega_{\mathrm{D}} \approx \num{0.89} \, \omega_x$ shown in the inset indicate a weak excitation of the center-of-mass mode.
	(b) Time evolution of the density profile~$n(x)$ in the stripe phase for the same scenario as in (a), explicitly revealing the excitation of the translational motion of the stripes. In the linear regime, the velocity of the stripes is proportional to the perturbation strength~$\lambda$ according to $v(\lambda) \approx \num{0.48} \, \lambda \, \hbar k_0 / m$.%
}
\end{figure}

In \cref{fig:beating_dispersion_symmetric:dispersion}, we also report the dispersion of the center-of-mass (dipole) mode, which is excited by suddenly removing a perturbation proportional to the operator~$x$, corresponding to a  shift of the harmonic trap along the $x$~direction. Above $\omegacritical$, the center-of-mass operator~$x$ and the spin operator~$\sigma_z$ excite the same mode, similar to the case of the operators $x^2$ and $x \sigma_z$ discussed above. Both the breathing and the dipole frequencies decrease when approaching the transition to the single-minimum phase, where the effective mass increases, inducing sizable non-linear effects~\cite{Li2012,Zhang2012}.

At the transition to the supersolid phase, both the breathing and the dipole frequencies exhibit a small jump, reflecting the first-order nature of the supersolid--superfluid transition. Entering the supersolid phase, one expects the emergence of the Goldstone mode that corresponds, in uniform matter, to the rigid translation of stripes. In a harmonic trap, the frequency of this motion is not exactly zero, but still much smaller than the oscillator frequency~$\omega_x$. The existence of this \enquote{zero-frequency} Goldstone mode can be inferred employing a sum-rule argument, according to which a rigorous upper bound to the lowest-energy mode excited by the operator~$x$ is given by $\omega_{\mathrm{lowest}} \le \omega_x / \sqrt{1 + 2 \recoilenergy \chi}$, where $\chi$ is the magnetic polarizability~\cite{Li2012}. This upper bound practically coincides with the center-of-mass frequency~$\omega_{\mathrm{D}}$ if $\Omega > \omegacritical$, while below $\omegacritical$ the calculated value of $\omega_{\mathrm{D}}$ violates the bound due to the large value of $\chi$, revealing the existence of a new low-frequency mode.

To shed light on the nature of this low-energy excitation, we apply a uniform spin perturbation proportional to the operator $\sigma_z$, causing a magnetic polarization of the system. After removing the perturbation~\footnote{To avoid the excitation of high-frequency modes on the order of the Raman coupling, we switch off the perturbation smoothly within a time interval~$\tau$ chosen such that $2 \pi \hbar / \Omega \ll \tau \ll 2 \pi / \omega_x$.}, one would expect the polarization to oscillate around its equilibrium value, driven by the Raman coupling.
Indeed, above $\omegacritical$, after a short initial decrease reflecting the contribution of the high-frequency gapped spin branch to the static magnetic polarizability~\cite{Martone2012}, the polarization oscillates at the center-of-mass frequency [see \cref{fig:zero_frequency_goldstone:polarization}]. The fact that the spin operator excites the center-of-mass mode is a manifestation of the hybridization mechanism between the operators $x$ and $\sigma_z$, similar to the one discussed above for $x^2$ and $x \sigma_z$.

In the stripe phase, we find instead that the polarization remains locked to its initial value throughout the simulation time, with a residual small-amplitude oscillation stemming from a weak excitation of the center-of-mass mode by the spin operator~$\sigma_z$. We have verified that the locking of the polarization survives longer than $1000$ times the oscillator time~$2 \pi / \omega_x$, confirming the anticipated low frequency of the Goldstone mode.
Remarkably, releasing the spin perturbation has the effect of applying a boost to the stripes, causing their translation at a constant velocity proportional to the perturbation strength, as illustrated in \cref{fig:zero_frequency_goldstone:density}. The translation of the stripes is practically independent of the center-of-mass motion, thereby providing direct evidence for the excitation of the zero-frequency Goldstone mode.
By contrast, after suddenly shifting the trap center, corresponding to a perturbation by the dipole operator $x$, the center of mass oscillates around equilibrium both in the superfluid and in the supersolid phase (not shown). This shows that the zero-frequency Goldstone mode contributes only marginally to the static dipole polarizability, whereas its strong excitation by the operator~$\sigma_z$ implies that it constitutes the predominant contribution to the magnetic polarizability in the stripe phase.

\begin{figure}
\includegraphics[width=\columnwidth]{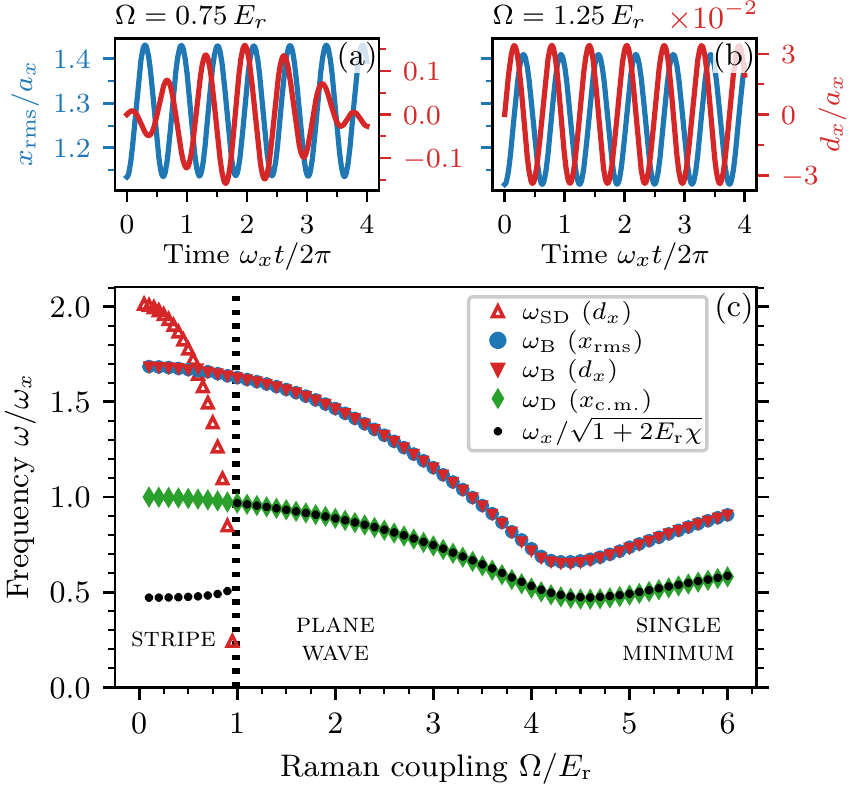}%
\subfloat{\label{fig:beating_dispersion_asymmetric:stripe}}%
\subfloat{\label{fig:beating_dispersion_asymmetric:pw}}%
\subfloat{\label{fig:beating_dispersion_asymmetric:dispersion}}%
\caption{%
	\label{fig:beating_dispersion_asymmetric}Same as \protect\cref{fig:beating_dispersion_symmetric}, but for strongly asymmetric intraspecies interactions, as relevant for $^{39}$K. The asymmetry leads to a smooth crossover between the plane-wave and the single-minimum regime. Nevertheless, in the stripe phase, the beating effects (a) and the additional frequencies (c) characteristic of the Goldstone mode are evident, while they are absent in the plane-wave phase (b).%
}
\end{figure}

In the last part of this Letter, we focus on a configuration characterized by strongly asymmetric intraspecies interactions, $g_{\upcomponent \upcomponent} \gg g_{\downcomponent \downcomponent} \approx g_{\upcomponent \downcomponent}$, yielding $g_{\densitycomponent \densitycomponent} \approx g_{\spincomponent \spincomponent} \approx g_{\densitycomponent \spincomponent}$.
The main motivation is to explore the consequences of the high spin polarization on the stripe phase, relevant for the case of $^{39}$K, which has recently become available for experiments on spin--orbit-coupled Bose--Einstein condensates~\cite{Tarruell2020}. We consider a set of scattering lengths given by $a_{\upcomponent \upcomponent} = \SI{252.7}{\bohr}$, $a_{\downcomponent \downcomponent} = \SI{1.3}{\bohr}$, and $a_{\upcomponent \downcomponent} = \SI{-6.3}{\bohr}$, realizable in $^{39}$K by using Feshbach resonances near a magnetic field of $B \approx \SI{389}{\gauss}$~\cite{Tarruell2020}. These values are consistent with the stability condition $g_{\upcomponent \upcomponent} g_{\downcomponent \downcomponent} > g_{\upcomponent \downcomponent}^2$.
Furthermore, we choose $k_0 = 2 \pi / \lambda_{\mathrm{Raman}}$ with $\lambda_{\mathrm{Raman}} = \SI{768.97}{\nano\meter}$~\cite{Tarruell2020}, $(\omega_x, \omega_y, \omega_z) = 2\pi \times (\num{50}, \num{200}, \num{200}) \, \si{\hertz}$, $N = \num{e5}$, and $\delta = 0$.
With these parameters, the transition to the supersolid phase occurs at $\omegacritical \approx \num{1.0} \, \recoilenergy$, leading to fringes with high contrast and hence to observable supersolid effects.

Because of the strong asymmetry in the interspecies interactions ($g_{\upcomponent \upcomponent} / g_{\downcomponent \downcomponent} \approx 194$), energetic minimization favors a large value of the spin polarization. In particular, the polarization is  large also in the stripe phase, as opposed to the symmetric case, where it instead vanishes. A simple estimate of this effect can be obtained in uniform matter, where, for $\Omega \to 0$, the variation of the energy~\labelcref{E} yields the expression $s_z/n = -g_{\densitycomponent \spincomponent} / g_{\spincomponent \spincomponent}$, which amounts to a polarization of $\braket{\sigma_z} \approx \num{-0.94}$ for our parameters. An important consequence of the strong asymmetry of the intraspecies interactions is that the central density, dominated by the majority component~$\Psi_\downcomponent$, is strongly enhanced with respect to the usual values characterizing symmetric configurations. This effect is accompanied by a shrinking of the cloud radius and the occurrence of fringes with high contrast in the minority component~$\Psi_\upcomponent$.

Following the procedure employed in the symmetric case, we consider the relative displacement~$d_x$ of the two components after a sudden quench of the trapping frequency [see \cref{fig:beating_dispersion_asymmetric:stripe,fig:beating_dispersion_asymmetric:pw}].
Also in this case, we observe a clear beating effect, revealing the occurrence of a Goldstone mode of spin nature in the stripe phase.
Because of the large polarization of the system, this mode mainly corresponds to the motion of the minority component, while the majority component remains practically at rest.
 
In \cref{fig:beating_dispersion_asymmetric:dispersion}, we show the dispersion law of the resulting elementary excitations as a function of $\Omega$, along with the dispersion of the center-of-mass mode excited by shifting the trap center.
At $\Omega \to 0$, the spin-dipole frequency is larger than the center-of-mass frequency as a result of the negative interspecies interaction~$g_{\upcomponent \downcomponent}$, in agreement with \cref{omegaSD}.
Because of the asymmetry of the intraspecies interactions, the transition between the plane-wave and the single-minimum phase is less sharp than in the symmetric case and actually becomes a smooth crossover. Furthermore, we find that the contrast of the stripes does not exhibit a jump at the supersolid--superfluid transition, but vanishes continuously, which may indicate that, due to the smallness of $g_{\downcomponent \downcomponent}$, the system is still rather far from the thermodynamic limit.
Nonetheless, the qualitative features of the excitation spectrum, including the occurrence of the zero-frequency Goldstone mode, are similar to the symmetric case.

In conclusion, we have provided accessible signatures of the Goldstone modes exhibited by the stripe phase of a harmonically trapped spin--orbit-coupled Bose--Einstein condensate. The Goldstone modes are revealed by a characteristic beating effect in the spin-dipole observable, following an experimentally straightforward density perturbation, and by the dynamic excitation of the translational motion of the stripes (zero-frequency Goldstone mode), following the release of a uniform spin perturbation.
Both configurations with symmetric and highly asymmetric intraspecies interactions provide promising platforms for observing these effects.
Our results illustrate the rich hybridization phenomena that appear in a supersolid system with non-trivial spin degree of freedom and how these open new routes to identifying fundamental hallmarks of supersolidity.

\appendix

\begin{acknowledgments}
We thank Li Chen, Han Pu, Jean Dalibard, Gabriele Ferrari, and the ICFO team led by Leticia Tarruell for fruitful discussions.
This project has received funding from the European Research Council (ERC) under the European Union’s Horizon 2020 research and innovation programme (ERC StG StrEnQTh, Grant Agreement No.\ 804305), the Provincia Autonoma di Trento, and Q@TN --- Quantum Science and Technology in Trento.
The authors acknowledge support by the state of Baden-Württemberg through bwHPC.
\end{acknowledgments}

\bibliography{references}
 
\end{document}